
\input harvmac

\def\a{\alpha}   \def\G{\Gamma}
\def\d{\delta}  \def\ee{\epsilon} 
   
 \def\l{\lambda} \def\L{\Lambda}  
    
\def\s{\sigma}   
    
\def\Ps{\Psi}  \def\O{\Omega} \def\vf{\varphi}
\def\pa{\partial}   \def\half{{1\over
2}} \def\dx{d^3\!x\ }

\Title{}{Yang-Mills Fields and Riemannian Geometry\footnote{$^*$} {This
work is supported in part by funds provided by the U.S.  Department of
Energy (D.O.E.) under cooperative agreement \#DE-FC02-94ER40818 and by
the M.E.C. (Spain).}}

\centerline{Peter E. Haagensen} \centerline{ {\it Departament
d'Estructura i Constituents de la Mat\`eria}} \centerline{ {\it Facultat
de F\'\i sica, Universitat de Barcelona}} \centerline{ {\it Diagonal,
647~~~08028 Barcelona, SPAIN}}\medskip \centerline{ {\it and}}\medskip
\centerline{Kenneth Johnson} \centerline{{\it Center for
Theoretical Physics, Laboratory for Nuclear Science}} \centerline{ {\it
Massachusetts Institute of Technology, Cambridge, MA~~~02139, USA}}

\bigskip

\centerline{Abstract}\bigskip

It is possible to define new, gauge invariant variables in the Hilbert
space of Yang-Mills theories which manifestly implement Gauss' law on
physical states.  These variables have furthermore a geometrical
meaning, and allow one to uncover further constraints physical states
must satisfy.  For gauge group $SU(2)$, the underlying geometry is
Riemannian and based on the group $GL(3)$.  The formalism allows also
for the inclusion of static color sources and the extension to gauge
groups $SU(N>2)$, both of which are discussed here.

\vfill

\bigskip
\noindent CTP\#2351

\noindent UB-ECM-PF 94/21

\Date{08/94}

\newsec{Introduction}\bigskip

The property of gauge invariance is at the heart of physical
manifestations and theoretical aspects of QCD as the theory of strong
interactions.  While in the perturbative regime it is intimately tied to
the renormalizability of the theory, in the strong coupling regime it is
an important and restrictive constraint on the ground state of the
theory, which in turn is intimately tied to the problem of confinement.
Naturally, in pure Yang-Mills theory, the same holds true.  There, a
first striking observation regarding the gauge symmetry is that if one
considers a theory with semisimple gauge group $G$ in the canonical
formalism and in the Weyl gauge $A_0^a=0$, while the basic variables
$A_i^a$ are $3 \dim G$ in number, there are only $2\dim G$ gauge
invariant degrees of freedom.  Gauss' law is of course what enforces
this, and one may then wonder whether there is a more appropriate set of
variables that puts in evidence this cyclic nature of the gauge
non-invariant part of $A_i^a$ and thus obviates Gauss' law, which in
these original variables are, to say the least, a complicated technical
nuisance to implement.  In the abelian theory, for instance, this is
straightforward because Gauss' law there simply states that the
longitudinal component of $A_i$ is cyclic.  Gauge invariance is
implemented simply by considering all states to depend only on the
transversal components of $A_i$, and that is basically the end of the
story.  To be sure, we can in no way consider this as a paradigm to
follow closely in the nonabelian case because, after all, the abelian
theory is a theory of free photons, while the nonabelian theory is a
complicated interacting theory.  Nonetheless, the hope still remains
that the use of local gauge invariant variables to describe the physical
space of the theory may in some way present itself as a viable and
fruitful procedure to follow.  This was the first and underlying idea
motivating this work.

This idea in itself is not new.  Goldstone and Jackiw \ref\gj{J.
Goldstone and R. Jackiw, {\it Phys.  Lett.} {\bf 74B} (1978) 81.} have
considered, for a pure Yang-Mills theory with gauge group $SU(2)$, a
realization of the canonical commutators in which the electric field
$E^{ai}$ rather than the vector potential is diagonal.  Unlike $A_i^a$,
the electric field is gauge covariant, and separating a gauge invariant
part from gauge angles in such a quantity is a much simpler matter.
Recently, a different proposal to use gauge invariant variables in the
electric formalism has also been presented \ref\bfh{M.  Bauer, D.Z.
Freedman and P.E.  Haagensen, {\sl ``Spatial Geometry of Electric Field
Representation of Non-Abelian Gauge Theories''}, preprint CERN-TH.
7238/94, to be published in {\it Nucl.  Phys.} {\bf B}.  Earlier work on
the electric field representation can also be found in A.G.  Izergin,
V.E.  Korepin, M.A. Semenov-Tian Shianskii and L.D.  Faddeev, {\it Teor.
Mat. Phys.} {\bf 38} (1979) 1.}, wherein the geometrical character of
the gauge invariant variables has been exploited in a similar fashion as
presented in this work.  The electric field formalism, however, quickly
becomes rather cumbersome, due to the fact that the magnetic energy
density in the Hamiltonian involves terms with up to four functional
derivatives.  We choose here to stay within the usual realization of the
canonical commutators.

Within this realization one may consider, as a first attempt, a
canonical change of variables from $A_i^a$ to $B^{ai}$, the color
magnetic field \ref\fhjl{K.  Johnson,{\sl ``The Yang-Mills Ground
State''}, in {\it QCD -- 20 Years Later}, Aachen, June 1992, Vol.  I, p.
795, eds.  P.M.  Zerwas, H.A.  Kastrup\semi D.Z.  Freedman, P.E.
Haagensen, K. Johnson and J.I.  Latorre, {\sl ``The Hidden Spatial
Geometry of Non-Abelian Gauge Theories''}, preprint CERN/TH. 7010/93,
CTP\#2238.  The idea of separating a gauge invariant part in the vector
potential is also in Y.A.  Simonov, {\it Sov.  J. Nucl.  Phys.} {\bf 41}
(1985) 835.}.  Locally, both have the same number of components, but the
difference is again that $B^{ai}$ transforms covariantly under gauge
transformations.  Building gauge invariant variables from it is as easy
as contracting it with itself in color: $\vf^{ij}\equiv B^{ai}B^{aj}$.
For $SU(2)$, there are six gauge invariant degrees of freedom at each
point in space, and there are just as many components in the symmetric
$3\times 3$ tensor $\vf^{ij}$.  For larger groups there are more gauge
invariant degrees of freedom ($2\dim G$), but there are also higher
invariant tensors (such as $d^{abc}$ for $SU(3)$), and contraction of
$B^{ai}$'s with these complete the number of gauge invariant variables
needed.  In terms of $B^{ai}$, Gauss' law is simply the generator of
rotations in color space, and in terms of gauge invariant variables it
is nothing, that is to say, it only contains variations w.r.t.~gauge
angles, and does not involve changes in gauge invariant variables at
all.  States $\Ps[\vf^{ij} ,\ldots]$ depending only on these gauge
invariant variables manifestly satisfy Gauss' law.  At this point, it
seems we are essentially done with our programme.  One question remains,
however, whose answer will lead to the unraveling of this first and most
direct attempt at solving Gauss' law: are we spanning {\it enough} of
the physical Hilbert space of the theory with these variables?
Certainly, we cannot assert that these variables span, say, Wilson
loops, which are gauge invariant variables in their own right, but are
we at least spanning {\it local} gauge invariant quantities properly?

Unfortunately, the answer to even this question is no.  The problem lies
in the fact that gauge fields suffer from Wu-Yang ambiguities
\ref\wy{T.T.  Wu and C.N.  Yang, {\it Phys.  Rev.} {\bf D12} (1975)
3845.  See also D.Z.  Freedman and R.R.  Khuri, {\it Phys.  Lett.} {\bf
329B} (1994) 263, and references therein.}: to one given configuration
of the magnetic field $B^{ai}$, it turns out there may correspond in
general many (or infinitely many) vector potential configurations
$A_i^a$.  That is to say, the map $A\to B$ is in general many-to-one.
For a given magnetic configuration admitting Wu-Yang ambiguous
potentials, the associated variables $\vf^{ij}$ are obviously
insensitive to the ambiguity, while other gauge invariant quantities,
say, $B^{ai}D_jB^{ak}$ or $D_iB^{aj}D_kB^{a\ell}$, are not.  There is no
way to represent these ``Wu-Yang sensitive" terms in the Hamiltonian in
terms of variables which are ``Wu-Yang insensitive".  In the original
path integral, Wu-Yang related potentials must be integrated over, since
they are not gauge related, while the integration over variables such as
$\vf^{ij}$ always misses these configurations.

This leads us to discard this approach, and instead turn to one we
consider more fruitful, and ultimately more elegant.  With the objective
of manifest gauge invariance still in mind, but also conscious of the
problems described above, we will seek to find new variables $u^a_i$
which again are gauge covariant, but now avoid Wu-Yang ambiguities
entirely.  These new variables will be defined through a set of first
order differential equations or, alternatively, through a variational
principle.  The absence of Wu-Yang ambiguities is manifestly seen from
the fact that an explicit expression for $A_i^a$ as a function of
$u^a_i$, $A[u]$, follows from these equations.  It will also be true, on
the other hand, that to a given configuration of the vector potential
there may correspond in general many $u^a_i$ configurations.  As an
instance of this, we will see that the linearity in $u$ of the
transformation of variables causes the scale of $u$ to be left
undetermined.

At the outset we will point out and explore a symmetry enjoyed by both
the canonical variables of Yang-Mills theory and the Gauss law
generator.  That is the symmetry under $GL(3)$ reparametrizations, which
will turn out to be very natural in the sense that it does not
necessitate the introduction of a metric or diffeomorphism covariant
quantities.  Certainly, the Hamiltonian will not possess this symmetry,
as we know Yang-Mills theory is after all {\it not} generally covariant.
Yet, the manifest violation of this symmetry only happens at the level
of the Hamiltonian, and we will be able to use the $GL(3)$ covariance as
a guiding principle throughout.  For the case of gauge group $SU(2)$,
the new variables will play the role of a triad or dreibein, and a
metric is thus introduced which will turn out to be precisely the gauge
invariant variables appropriate to the problem.  A natural relationship
will be seen to emerge between gauge invariant objects and the
Riemannian geometry associated to this metric.  Apart from the manifest
breaking term mentioned above, the Hamiltonian will be built entirely
out of geometrical objects.  The Gauss law and gauge Bianchi identities
will also have a straightforward translation into their geometric
analogues.  Also quite naturally, our change of variables will imply
that pure gauge configurations will correspond to flat metrics and
vice-versa.

In terms of the new gauge invariant and geometric variables, the
Hamiltonian will turn out to be nonlocal.  We shall also see that in our
geometrical formalism the requirement of finite energy will lead to both
global and local constraints on physical wave functionals.  The former
arise in connection with the zero modes of the operator
$\epsilon^{ijk}D_j$, while the latter appear in singular regions where
the Einstein tensor built out of our metric variables is not invertible.
These constraints are a welcome sign that our approach is indeed
bringing out truly nonperturbative aspects of the theory, not accessible
to perturbation theory.

The plan of the paper is as follows.  In order to set notation and
conventions, in Sec. 2 we briefly review the canonical formalism for
Yang-Mills theories.  In Sec. 3, we exploit the $GL(3)$ symmetry in
order to introduce the geometrical setting and variables to be used.  We
pursue this further in Sec. 4, where the geometrical translation of
Yang-Mills formulas is presented, and physical consequences are
analyzed.  A discussion on the inclusion of static color sources, which
modify Gauss' law, is presented in Sec. 5. Finally, in Sec. 6, we
briefly outline the extension of our work to gauge group $SU(N>2)$.
\bigskip

\newsec{The $GL(3)$ Properties of Yang-Mills Canonical Variables}
\bigskip

In this section it shall be shown that in the Hamiltonian formulation of
the Yang-Mills gauge field theory there is a symmetry of the canonical
variables under general coordinate transformations in 3-dimensional
space.  The group is $GL(3)$, where the transformations are
$x'^i=x'^i(x)$ and contravariant tensor indices $j$ transform with the
matrix: \eqn\gct{ {\partial x'^i\over\partial x^j}\ ,} while covariant
tensor indices transform with the inverse matrix.  To show this, the
equations which define the Hamiltonian version of the field theory will
be reviewed.  One should keep in mind that eventually the quantum
mechanics will be formulated in the Schr\"odinger picture.  The
Hamiltonian is \eqn\ham{H\ =\ \half\ \int \dx \left({B^{ak}B^{ak}\over
g_s^2}+ g_s^2E^{ak}E^{ak}\right)\ .} The color magnetic field, $B^{ak}$,
which appears above is defined to be a function of the color vector
potential $A^a_i$, which is the canonical coordinate: \eqn\bofa{B^{ak}\
=\ \ee^{kij}F^a_{ij}\ =\ \ee^{kij} (\pa_i A^a_j+\half\ f^{abc}A^b_i
A^c_j)\ .} The canonical momentum is the color electric field $E^{ak}$,
where \eqn\comm{[\ A^a_i(x),E^{bj}(y)\ ]\ =\ i\d^{ab}\d^j_i\d (x-y)\ .}
The Hamiltonian $H$ is locally gauge invariant: \eqn\trham{[\ H,{\cal
G}^a(x)\ ]\ =\ 0\ ,} where \eqn\gauss{{\cal G}^a(x)\ =\ D_i E^{ai}(x)\
\equiv\ \pa_i E^{ai}(x)+ f^{abc}A^b_i(x)E^{ai}(x)\ ,} and the generator
${\cal G}^a$ produces the infinitesimal local $SU(N)$ transformations of
the canonical variables, \eqn\trAE{\eqalign{[\ A^a_i(x),{\cal G}^b(y)\
]\ &=\ -i\d^{ab}\pa_i\d (x-y)\ +if^{abc}A^c_i(x)\d (x-y)\cr [\
E^{ai}(x),{\cal G}^b(y)\ ]\ &=\ if^{abc}E^{ci}(x)\d (x-y)\ .}} The
generators form a local algebra \eqn\gg{ [\ {\cal G}^a(x),{\cal G}^b(y)\
]\ =\ if^{abc}{\cal G}^c(x)\d (x-y)} of a local compact group.  Under
finite $SU(N)$ transformations the canonical variables transform as
follows: \eqn\finA{\O^{-1}A^a_i(x)\O\ =\ U^{ab}(x)A^b_i(x)-{1\over N}
f^{abc}U^{bb'}(x)\pa_i U^{cb'}(x)} \eqn\finE{\O^{-1}E^{ai}(x)\O\ =\
U^{ab}(x)E^{bi}(x)\ .} Equations \bofa,\comm\ and \gauss-\finE\ are not
only invariant under the group of $SU(N)$ gauge transformations, they
are also invariant under the group $GL(3)$ of general coordinate
transformations.  This is true without the intervention of a space
metric.  The only requirements are that $A^a_i$ transform as a covariant
vector and $E^{ai}$ as a contravariant vector density.  In this case the
ordinary space derivatives which appear are equivalent to covariant
derivatives.  It is for this reason that the lower and upper index
notation on $A$ and $E$ has been used.  It otherwise has no
significance.  One also notes that $B$ transforms as a contravariant
vector density as a consequence of \bofa\ and the tensor property of
$A$.  However, the Hamiltonian itself has no simple property under this
group except for the standard global symmetries of spatial rotations and
translations, which are but a tiny subgroup of $GL(3)$.  Because the
Hamiltonian has no simple property under this group, there is no very
simple analog of this $GL(3)$ symmetry of the canonical variables in a
Lagrangian formulation of the theory.\bigskip

\newsec{The Introduction of a Metric}

To begin, the development will be limited to the simplest case of
$SU(2)$.  The extension to $SU(N>2)$ will be outlined in Sec. 6. For
$SU(2)$ the structure constants are $f^{abc}=\ee^{abc}$.  A new
coordinate variable $u^a_i$ to replace the vector potential will now be
defined.  It will transform covariantly under $SU(2)$ rather than as a
gauge connection, \finA . The guide in our effort to find such a
coordinate will be the condition that the $GL(3)$ symmetry of the
original phase space of $A$ and $E$ be maintained.  Thus, under $GL(3)$,
$u^a_i$ will be a covariant vector just as the vector potential.  Both
$A$ and $u$ have a total of nine components at each point in space so
nine equations will be provided to define the relationship between them.
Since under local gauge transformations $u^a_i$ should transform as a
vector, \eqn\finu{\O^{-1}u^a_i(x)\O\ =\ U^{ab}(x)u^b_i(x)\ .} we define
the transformation between $u$ and $A$ so that \finu\ be consistent with
\finA , which is assured if \eqn\defu{\ee^{ijk}(\pa_j
u^a_k+\ee^{abc}A^b_j u^c_k)\ =\ 0\ .} Because of the tensorial property
of the ordinary curl of a covariant vector, it is clear that this
equation is also covariant under $GL(3)$ transformations.  Further,
\defu\ is a set of nine first order linear differential equations to
relate the vector functions $u^a_i$ to the vector fields $A^a_i$.
Because the equations are linear and homogeneous, the global scale of
the coordinate $u$ is not determined by $A$.  When $u$ replaces $A$, the
functionals of $u$ will be constrained to be globally homogeneous.  The
mutual consistency of the equations for $u$ for a given $A$ is made
easier to understand by the fact that they can be obtained from a
variational principle \ref\hn{This same functional has appeared in the
first reference in \bfh\ and, in a different context, in M. Henneaux,
J.E.  Nelson and C. Schonblad, {\it Phys. Rev.} {\bf D39} (1989) 434,
and H. Nicolai and H.-J. Matschull, {\it J. Geom. and Phys.} {\bf 11}
(1993) 15.}. Define a functional $W[u]$, \eqn\w{W[u]\
=\ -\half\int\dx \ee^{ijk}u^a_i(x)\pa_j u^a_k(x)\ .} $W[u]$ is a global
$GL(3)$ invariant functional of $u$.  Further, \eqn\finw{ W[Uu]\ =\
W[u]+\half\int\dx \ee^{abc}\ U^{b'b}\pa_j U^{b'c}\ u^{aj}\det u\ ,}
which is established using the identity \eqn\uinv{\ee^{ijk}u^a_i u^c_k\
=\ \ee^{abc}u^{bj}\det u\ ,} where $u^{bj}$ is the inverse of the matrix
$u^a_i$, \eqn\uuinv{u^a_i u^{bi}\ =\ \d^{ab}\ ,} and $\det u$ stands for
the determinant of the matrix $u^a_j$.  It then follows from \finw\ that
the quantity \eqn\dwdu{A^a_i\ \equiv\ {\d W\over\d (u^{ai}\det u)}}
transforms as a gauge connection \finA\ when $u$ transforms covariantly
as in \finu . It is straightforward to work out an explicit expression
for the quantity $A$ and obtain \eqn\Aofu{A^a_i\ =\
{(\ee^{nmk}\pa_mu^b_k)(u^a_nu^b_i-\half u^b_nu^a_i) \over \det u}\ .} It
then can be checked that this formula is equivalent to the original
defining equation for $u$, \defu .

So far all of the equations are $GL(3)$ covariant without the appearance
of a metric tensor, but one can now see that a metric tensor has
implicitly been introduced by the definition \defu\ of $u$.  It follows
from \defu\ that \eqn\covu{\pa_j u^a_k+\ee^{abc}A^b_j u^c_k\ =\ \left\{
\matrix{s \cr jk \cr}\right\}\ u^a_s\ ,} where the curly bracket
quantity is symmetric in the indices $j,k$.  Further, since the left
hand side of \covu\ transforms as a vector under $SU(2)$, the curly
bracket is an $SU(2)$ invariant.  If \covu\ is multiplied by $u^a_m$ and
one forms the symmetric part of the resulting equation in $k,m$ the
result is \eqn\compat{\pa_j(u^a_mu^a_k)-\left\{ \matrix{s \cr jk
\cr}\right\}\ u^a_s u^a_m-\left\{ \matrix{s \cr jm \cr}\right\}\ u^a_s
u^a_k\ =\ 0\ .} When the curly brackets are symmetric in the lower
indices, they are given by the unique and well-known solution
\eqn\christoffel{ \left\{ \matrix{i \cr jk \cr}\right\}\ =\ \half\
g^{im}(\pa_j g_{mk}+\pa_kg_{jm}-\pa_mg_{jk})\ ,} where
\eqn\metric{g_{ij}\ =\ u^a_i u^a_j\ ,} which makes it the symmetric
affine connection for the metric $g_{ij}$.  Thus \defu\ has implicitly
introduced a Riemannian geometry with a metric tensor which is a
function of $u$ and therefore of $A$.  The metric tensor is manifestly
gauge invariant.  One can now write equation \defu\ in either of two
equivalent forms.  Both express the invariance of $u$ under space
translations.  One equation expresses the invariance with respect to the
spatial geometry, while the other expresses the invariance with respect
to the gauge choice.  Here a notation for the geometric covariant
derivative of a vector is introduced, \eqn\geomd{\nabla_j u^a_k\ =\
\pa_j u^a_k- \left\{ \matrix{s \cr jk \cr}\right\}\ u^a_s\ ,} while the
$SU(2)$ gauge covariant derivative is, \eqn\geomd{D_j u^a_k\ =\ \pa_j
u^a_k+\ee^{abc}A^b_j u^c_k\ .} The invariance equations are
\eqn\inva{\nabla_j u^a_k+\ee^{abc}A^b_j u^c_k\ =\ 0\ ,} or equivalently,
\eqn\invb{D_j u^a_k- \left\{ \matrix{s \cr jk \cr}\right\}\ u^a_s\ =\ 0\
.} One should not fail to note that $\ee^{abc}A^b_i$ plays the role of a
spin connection, while $u$ is known as a dreibein or triad.  Here there
is the added feature that this spin connection may be regarded as an
$SU(2)$ Yang-Mills gauge potential.  Equation \inva\ can also be written
\eqn\aaa{A_i^a\ =\ -{1\over 2} \epsilon^{abc} u^{bj} \nabla_i u_j^c\ .}
In this form, it is made clear that when the metric has no curvature,
one can make a general coordinate transformation which makes $g_{ij}$
proportional to $\delta_{ij}$ and the covariant derivative equivalent to
an ordinary derivative in which case \aaa\ expresses the vector
potential in the form of a ``pure gauge'' where $u_j^a$ is a unitary
matrix.  Locally constant unitary matrices yield the zero potentials.
One can see from these considerations that the map $u=u(A)$ is in
general not unique.  However, because $u$ is a zero mode of a linear
operator, this non-uniqueness is not as troublesome as that of the
nonlinear operator giving the magnetic field $B[A]$, Eq.~\bofa . The
space of $u$ is larger than that of $A$, and we assume without giving a
proof that the space of $u$ includes all $A$.  When one goes the other
way, $A=A(u)$, there is a unique $A$ associated with each $u$ when $\det
u\neq 0$.  This follows from \defu.  If one $u$ were to give different
$A$'s, say $A$ and $\tilde{A}$, then
\eqn\bbb{\epsilon^{ijk}\epsilon^{abc}\left(A_j^b-\tilde{A}_j^b\right)
u_k^c=0\ .} The matrix $m^{ai,bj}=\epsilon^{abc}\epsilon^{ijk}u_k^c$ is
invertible if $\det u\neq 0$.  Hence, in this case $A=\tilde{A}$.

To summarize, the map $A=A(u)$ represents a function of $u$, while the
inverse $u=u(A)$ is not a function since it is one to many in degenerate
cases, which include those $g$ which correspond to pure gauges.

\newsec{Yang Mills Tensors in Terms of Geometric Tensors}

Starting with \invb, one may use the gauge Ricci identity
\eqn\ccc{[D_i,D_j]^{ac}\ =\ \epsilon^{abc} F_{ij}^b} to find that
\eqn\ddd{\epsilon^{abc} F_{ij}^b u_k^c\ =\ R^\ell_{\, kij} u_\ell^a}
where $R^\ell_{\, kij}$ is the Riemann tensor associated with the affine
connection \christoffel, that is \eqn\eee{R^\ell_{\,
kij}=\pa_i\G_{jk}^\ell-\pa_j\G_{ik}^\ell+ \G_{jk}^m\G_{im}^\ell-
\G_{ik}^m\G_{jm}^\ell\ ,} where the symbol $\Gamma^k_{ij}$ is used for
the affine connection for compactness of notation.  In three space
dimensions the Riemann curvature tensor can be written in the form
\eqn\fff{R_{\ell kij}\ =\ -{\epsilon_{\ell ku} \epsilon_{ijv}\over g}\
G^{uv} \ ,} where $g=\det g_{ij}$, and $G^{uv}$ is the Einstein tensor,
given by $$ G_{uv}\ =\ R_{uv}-{1\over 2} g_{uv} R, $$ with $R_{uv}$ the
Ricci curvature defined by $R_{\, usv}^s \equiv R_{uv}$, and $R=R^v_v$
the Ricci scalar.  One can equivalently write \ddd\ in the form
\eqn\ggg{B^{ai}\ =\ \sqrt{g}\ u_j^a\ G^{ij} \ .} To be precise, by
$\sqrt{g}$ we mean $\det u$, which can in principle take on both
positive and negative values.  This notation is chosen simply for
clarity of formulas.  Thus, in geometric form, the magnetic field is
expanded in terms of the dreibein $u$ by the Einstein curvature tensor.
The gauge invariant tensor which gives the Yang-Mills magnetic energy
density is \eqn\hhh{B^{ai} B^{aj}\ =\ g\ g_{k\ell}\ G^{ik} G^{j\ell}\ .}
This gives the gauge invariant tensor in a manifestly gauge invariant
form in terms of the metric $g_{k\ell}$.

Gauge and geometric forms of the Bianchi identities can now be worked
out.  The density $D_i B^{ai}$ can be expressed using \ggg\ and \invb\
in the form \eqn\iii{D_i B^{ai}\ =\ \sqrt{g}\ (\nabla_i G^{ij})\ u_j^a\
.} From \iii\ then \eqn\jjj{D_i B^{ai} = 0\ \Rightarrow\ \nabla_i G^{ij}
= 0} and vice-versa.  Of course $\nabla_i G^{ij}=0$ is the familiar
geometric Bianchi identity.

One can now turn to a discussion of the ``electric'' tensor.  A gauge
invariant tensor operator $e^{ij}$ can be defined by
\eqn\kkk{{\delta\over \delta A_i^a}\ \equiv\ \sqrt{g}\ u_j^a e^{ij}\ .}
Since $\delta /\delta A_i^a$ transforms as a vector under local gauge
transformations, the operator $e^{ij}$ is gauge invariant, and plays the
same role for the ``electric'' field as $G^{ij}$ does for the magnetic
field.  However, in contrast to $G^{ij}$, the tensor operator $e^{ij}$
is not symmetric in general.  The factor $\sqrt{g}$ $(=\det u)$ has been
included so that $e^{ij}$ transforms as an ordinary tensor under $GL(3)$
transformations.  We assume that by definition $\Psi$ acts as a scalar
under $GL(3)$.  The operator ${\cal G}_a$ of local gauge transformations
\eqn\mmm{i {\cal G}_a \ =\ D_i \left({\delta\over \delta A_i^a} \right)}
can be evaluated in the same way as the Bianchi identity, so \eqn\nnn{i
{\cal G}^a \ =\ \sqrt{g}\ u_j^a (\nabla_i e^{ij})\ .} Thus, if $\Psi$ is
gauge invariant so that \eqn\ooo{{\cal G}^a \Psi\ =\ 0\ ,} then
\eqn\ppp{\nabla_i e^{ij} \Psi\ =\ 0} and vice versa, that is, if
$\nabla_i e^{ij}\Psi\ =\ 0$, then $\Psi$ is gauge invariant.  Finally
the ``electric'' gauge invariant Yang-Mills tensor is,
\eqn\qqq{{\delta\Psi\over \delta A_i^a} {\delta\Psi\over \delta A_j^a} \
=\ g\ g_{nm}\ (e^{in}\Psi)(e^{jm}\Psi)\ .} Eq. \qqq\ may be compared
with \eqn\rrr{B^{ai} B^{aj} \Psi^2\ =\ g\ g_{nm} G^{in} G^{jm} \Psi^2\
.} In connection with the global invariance of $\Psi$ under $g_{ij}\to
\lambda g_{ij}$, to be analyzed below, it is worthwhile noting that both
energy densities, Eqs. \qqq\ and \rrr , are also globally homogeneous in
$g_{ij}$.

In order to proceed one must express the functional dependence of
$\Psi(A)$ in terms of the vector variable $u$.  This can be done through
\aaa\ so \eqn\s{\delta A_i^a\ =\ -{1\over 2} \epsilon^{abc} \delta
(u^{bj} \nabla_i u_j^c)} After a short calculation, one finds that
\eqn\ttt{\delta A_i^a\ =\ {\epsilon^{nm\ell}\over 2\sqrt{g}} u_m^a
\left(\nabla_i (u_n^b \delta u_\ell^b) +\nabla_\ell (\delta g_{ni})
\right)} and therefore \eqn\uuu{\eqalign{ \d\Psi\ =&\ \int d^3\!x\
{\delta\Psi\over \delta A_i^a} \ \delta A_i^a \cr =&\ \int d^3\!x\
{\epsilon^{nm\ell}\over 2}\ \left( \nabla_i (u_n^b \delta u_\ell^b)
+\nabla_\ell (\delta g_{ni}) \right) e^i_{\ m}\Psi \ .\cr}} This
equation should allow one to determine the electric variable $e^i_{\
m}\Psi$ in terms of the dependence of the wave functional on the vector
variable $u$.

If the wave functional $\Psi$ depends on $u_i^a$ only through the
``metric'', that is, the gauge invariant composite $g_{ij}=u_i^a u_j^a$,
then \ppp\ follows directly from \uuu . The same follows in reverse
implication, that is, if \ppp\ holds, then from \uuu\ it follows that
the only functional dependence of $\Psi$ is on the metric.

Since gauge invariant functionals can be expressed in terms of $g_{ij}$,
Wilson loops in principle can be expressed in this way.  A first step
would be to express the ``gauge connection'' $A_i^a$ in terms of
$u_i^a$, and then one would express the dependence on the curve in terms
of the metric.

Since for later purposes it will be useful to have the form of the
electric operator where it acts on states which are not restricted to be
gauge invariant we shall not impose that condition immediately.  For
this general case it will be helpful to introduce the operator
\eqn\abab{{\tilde e}^i_{\ j}\equiv e^i_{\ j}- {1\over 2}\ \delta^i_j
e^s_{\ s}} where the inverse of \abab\ is \eqn\acac{e^i_{\ j}=\tilde
e^i_{\ j}-\delta_j^i\tilde e^s_{\ s} } After a short calculation one
finds that \eqn\adad{\epsilon^{inm} \nabla_n (\tilde e^j_{\ m})=
2{\delta\over \delta g_{ij}}+{\epsilon^{ijk}\over 2\sqrt g}\ u_k^a
\left(\epsilon^{abc} u_s^b {\delta\over\delta u_s^c} \right)} where as
already noted the second term is absent when the operator $e$ acts on
gauge invariant states.  It will be convenient to introduce a gauge
invariant operator \eqn\aeae{ {\cal F} _k \equiv i
u_k^a\left(\epsilon^{abc}u_s^\ell{\delta\over\delta u_s^c}\right) =
iu_k^a{\cal G}^a (x)} where ${\cal G}^a (x)$ is the gauge generator
expressed in terms of $u$.  In this case \adad\ takes the form
\eqn\afaf{\epsilon^{inm} \nabla_n (\tilde e_m^j) = 2 {\delta\over \delta
g_{ij}} + i{\epsilon^{ijk}\over 2\sqrt g}\ {\cal F}_k} The operator
${\cal F}_k$ obeys the commutation rules \eqn\agag{[{\cal F}_k(x), {\cal
F}_\ell(y)] = -i {\epsilon_{k\ell m}\over\sqrt g}\ {\cal F}^m
\delta(x-y)} and thus the ${\cal F}_k$'s may be regarded as a gauge
invariant set of operators which act in an ``intrinsic'' sense, to
separate the gauge dependent and gauge invariant parts of $\Psi$.  In
operator form \eqn\ahah{\left[{\cal F}_k(x) ,{\delta\over\delta
g_{ij}(y)}\right] = - \left[\delta_k^i {\cal F}^j + \delta_k^j {\cal
F}^i \right] \delta(x-y) \ .} Eqs. \afaf\ through \ahah\ may be regarded
as the complete ``polar'' decomposition of the operators of the electric
field.

It has already been noted that a consequence of these equations, if
$\Psi$ belongs to the gauge invariant subspace of the Hilbert space,
will be that its functional dependence will be solely on the metric
$g_{ij}$, \eqn\aiai{{\delta\Psi\over\delta g_{ij}} =
{\epsilon^{mni}\over 2} \nabla_m \tilde e^j_{\ n} \Psi} The expression
needed for $\tilde e^j_{\ n}\Psi$ and hence by means of \acac\ for
$e^j_{\ n}\Psi$, requires the inverse of the linear operator which
appears in \aiai.  This means, in general, that $e^j_{\ n}\Psi$ will
depend non-locally on ${\delta\Psi\over\delta g_{ij}(x)}$, and the
Hamiltonian will thus be a non-local functional.

To obtain a more explicit, albeit formal, expression for $\tilde{e}^j_{\
n}\Psi$, one must consider the eigenvalue problem for the operator to be
inverted: \eqn\top{T^i_{\ n}\phi_\alpha^{nj}\equiv {1\over\sqrt{g}}
\epsilon^{im}_{\ \ \ n}\nabla_m \phi_\alpha^{nj} =\lambda_\alpha
\phi_\alpha^{ij} \ .} Because $T$ is a real symmetric operator, the
tensor eigenfunctions $\phi_\alpha$ can be chosen to form a complete
orthonormal set.  Inversion will then be achieved through the formal
expression for the Green's function: \eqn\green{{\cal
H}^{im,jn}(x,y)={\sum_\alpha}^{'}
{1\over\lambda_\alpha}\phi_\alpha^{im}(x)\phi_\alpha^{jn}(y)\ .} In the
above, the prime on the sum means a sum only over states $\alpha$ for
which $\lambda_\alpha\neq 0$, that is to say, the inversion can only be
done in the subspace orthogonal to the zero modes of $T$.  As it turns
out, $T$ will always have at least one zero mode, given by the metric
itself, and possibly more for particular configurations.  In the
Appendix, we consider this question more carefully, and give arguments
as to why these zero modes should not affect the present discussion.

It is now straightforward to write the final expression for $\tilde{e}
^{jn}\Psi$: \eqn\finale{\tilde{e}^{jn}\Psi(x)=2\int d^3\!y\ {\cal
H}^{nj}_{\ \ rs} (x,y){\delta\Psi\over\delta g_{rs}(y)} \ .} From \acac\
and \kkk\ one then easily obtains an expression for the electric field
operator $\delta\Psi/\delta A^a_i$.

 From \aiai\ it is also possible to see that the requirement of
finiteness of $e^j_{\ n}\Psi$ will impose local constraints on gauge
invariant states.  For it follows from the geometric Ricci identity for
a mixed tensor \eqn\ajaj{[\nabla_i,\nabla_j]\sigma_n^m = \sigma_n^s R_{\
sij}^m -\sigma_s^m R_{\ nij}^s} applied to \aiai, that
\eqn\akak{\nabla_i \left({\delta\Psi\over\delta g_{ij}}\right) =
-{1\over 2} \epsilon^{jsu} (e_{sv}\Psi) G_u^v} This means that there are
local restrictions on $\Psi$ which follow from the requirement that
$(e_{sv}\Psi)$ remains finite when the curvature tensor vanishes at {\it
any\/} point in space.  When $G=0$ we can look at these restrictions
expressed in terms of coordinates where $g_{ij}=\delta_{ij}$ and
$\partial_k g_{ij}=0$, when they are $${\partial_i
\left({\delta\Psi\over\delta g_{ij}}\right)=0}$$ at the point where
$G^{ij}=0$, or $\Psi^2 B^{ai} B^{aj} =0$.  Thus restrictions are imposed
on {\it gauge invariant\/} functions for which the electric tensor
$(E^{ai}\Psi)(E^{aj}\Psi)$ is finite.  It is easy to show that such
requirements of finite $E$ go beyond simply $G=0$, if any two principal
components of the curvature vanish, then $$\nabla_i
\left({\delta\Psi\over\delta g_{iz}}\right)=0$$ where $z$ is the
principle direction of the component of curvature which is not zero.

To conclude this section, it should be pointed out that since it is the
canonical variables of the gauge field rather than the Hamiltonian which
have the simple $GL(3)$ tensorial properties, it has been this which
allowed the introduction of gauge invariant metrics.  Here the metric
which has been used is the one associated with the vector potential
viewed as a spin connection.  One could now use this metric to form a
new type of gauge theory with a Hamiltonian which is $GL(3)$ invariant,
namely, one formed by using the metric $g_{ij}$ and density $\sqrt{g}$
rather than the ``flat" metric $\d_{ij}$ used to form the Yang-Mills
Hamiltonian.  One might argue that at least on long scales these very
different Hamiltonians might have a significant dynamical similarity.

\bigskip\newsec{Extension to ``Static'' Color Sources}\bigskip

The Hilbert space $\Psi(u_i^a)$ is large enough to allow the inclusion,
along with the Yang--Mills field, of the coupling to ``static'' color
sources.  However, to do this one must use several copies,
\eqn\alal{\Psi_{a_1\ldots a_n} (u_i^a)} where the discrete labels
$a_1,a_2,\ldots a_n$ refer to $d_1 d_2 \ldots d_n$ copies of the Hilbert
space where $d_1,d_2,\ldots$ are the dimensions of representations of
the group SU(2).  The local gauge group is now defined to be
\eqn\amam{\Omega^{-1} \Psi_{a_1\ldots a_n} (u_i^a (x)) = U_{a_1
a_1'}^{(d_1)} (x_1) \ldots U_{a_n a_n'}^{(d_n)} (x_n) \cdot
\Psi_{a_1'\ldots a_n'} (U^{aa'} (x) u_i^{a'} (x))} where $U_{aa'}^{(d)}$
is the $d$-fold irreducible representation of the group SU(2).  The
``static'' positions of the sources are $x_1,x_2,\ldots x_n$.  The
infinitesimal generator takes the form, \eqn\anan{\bar {\cal G}^a(x) =
{\cal G}^a (x) + \sum_{\alpha=1}^n \Lambda_{(\alpha)}^a
\delta(x-x_\alpha)\ ,} where $\L^a_{(\a )}$ are the matrix generators of
$SU(2)$ in some representation.
The Yang--Mills Hamiltonian commutes with both parts
of \anan, but now the ``physical'' states are those where \eqn\aoao{\bar
{\cal G}^a (x) \Psi=0} or \eqn\apap{\bar {\cal F}_k \Psi =0} where as
above \eqn\aqaq{\bar {\cal F}_k = u_k^a (x) \bar {\cal G}^a (x) = {\cal
F}_k (x) + \sum_{\alpha=1}^n T_k^{(\alpha)} \delta(x-x_\alpha)} The
matrix operators are \eqn\arar{T_k^{(\alpha)} = u_k^a (x_\alpha)
\Lambda_{(\alpha)}^a} These obey the commutation rules
\eqn\asas{[T_k^{(\alpha)},T_\ell^{(\beta)}] = i {\epsilon_{k\ell
m}\over\sqrt{g_\alpha}}\ T^{(\alpha)m} \delta^{\alpha\beta}\ ,} where
$g_\a =\det g(x_\a )$, and
\eqn\atat{\left[{\delta\over\delta g_{ij} (x)},T_k^{(\alpha)}\right] =
\left(\delta_k^i T^{(\alpha)j} +\delta_k^j T^{(\alpha)i}\right)
\delta(x-x_\alpha)} The equation for the electric field is found in
combination with \apap, \aqaq\ and \afaf, \eqn\auau{\epsilon^{inm}
\nabla_n (\tilde e_m^j) \Psi = 2 {\delta\Psi\over\delta g_{ij}}
-i{\epsilon^{ijk}\over 2\sqrt g} \cdot \sum_{\alpha=1}^n T_k^{(\alpha)}
\delta (x-x_\alpha) \Psi} In addition the wave functionals in \auau\
obey the ``gauge'' restrictions, \eqn\avav{\left({\cal F}_k(x) +
\sum_{\alpha=1}^n T_k^{(\alpha)} \delta (x-x_\alpha) \right) \Psi = 0}
This means the parts of the space where ${\cal G}^a (x) \Psi \neq 0$ are
involved in \auau.  However, since \asas, \atat\ and \auau\ only involve
the variables $g_{ij}(x)$ and $T$, the explicit form of the restrictions
imposed by \avav\ will not enter as long as they can be consistently
applied.

\bigskip \newsec{Gauge Group $SU(N>2)$}\bigskip

The extension of the formalism to gauge groups $SU(N>2)$ can now be
outlined.  The main difficulty to be encountered is that the gauge group
here will not be the tangent space group of a 3-dimensional Riemannian
space, as was the case for gauge group $SU(2)$.  That will turn out to
require a new approach to the ``geometrization" of the gauge theory, and
although we do not present the entirely geometrical theory here, this
approach will be exemplified in some detail in the calculation of the
magnetic field for gauge group $SU(3)$.  The full geometrization of the
theory would be outside the scope of this article, and is left for
future work.

One begins by defining the functional $W[u]$ for $SU(N)$, which is
identical in form to \w : \eqn\wn{W[u]\ =\ -\half\int\dx
\ee^{ijk}u^a_i(x)\pa_j u^a_k(x)\ ,} with the difference that now $u^a_i$
are vectors of $SU(N)$, so that the index $a$ runs from 1 to $N^2-1$.
Local $SU(N)$ gauge transformations $u\rightarrow Uu$ lead to the
following variation in $W$: \eqn\finwn{ W[Uu]\ =\ W[u]-\half\int\dx
\ee^{ijk} u^{a'}_i u^{b'}_k\ U^{aa'}\pa_j U^{ab'}\ .} Because color
indices now belong to $SU(N>2)$, one has \eqn\euu{\ee^{ijk} u^{a'}_i
u^{b'}_k={2\over N}\ f^{a'c'b'}\bar{u}^{c'j}+ v^{j,a'b'}\ ,} where
$v^{j,a'b'}$ is a rank 2 antisymmetric tensor not in the adjoint (in
$SU(2)$, {\bf 3}$\times${\bf 3}= {\bf 1}+{\bf 3}+{\bf 5} and
antisymmetrization singles out the adjoint alone; in $SU(3)$, for
instance, {\bf 8}$\times${\bf 8}= {\bf 1}+{\bf 8}+{\bf 8}+{\bf
10}+${\bf\bar{10}}$+{\bf 27} and antisymmetrization singles out not only
the adjoint, but also the {\bf 10} and the ${\bf\bar{10}}$).  However,
one also has, for $U$ in $SU(N)$, \eqn\udu{U^{aa'}\pa_j
U^{ab'}=f^{a'b'c'}\pa_j\omega^{c'}\ ,} which belongs to the adjoint and
projects out $v^{a'b'}$.  Equations \euu\ and \udu\ then give
\eqn\ubar{\bar{u}^{ai}={1\over 2}\ f^{abc}\epsilon^{ijk}u^b_j u^c_k\ ,}
and \eqn\wuu{ W[Uu]\ =\ W[u]+{1\over N}\int\dx f^{abc} \bar{u}^{aj}\
U^{a'b}\pa_j U^{a'c}\ .} Therefore, again as in $SU(2)$, the quantity
\eqn\aia{A_i^a\equiv {\delta W\over\delta\bar{u}^{ai}}} transforms as a
gauge connection of $SU(N)$.  This is equivalent to defining $A_i^a$
through \eqn\defun{\ee^{ijk}(\pa_j u^a_k+f^{abc}A^b_j u^c_k)\ =\ 0\ .}

It is now convenient to define a basis for adjoint color vectors based
on the $O(3)$ subgroup of $SU(N)$, as follows:
\eqn\basissun{u^a_i,u^a_{\{ij\}},u^a_{\{ijk\}},\ldots} where
$\{\ldots\}$ means complete symmetry and tracelessness on any two
indices.  It is a quick combinatorial exercise to verify that for each
$a$ these basis vectors have respectively, $3,5,7,\ldots ,2N-1$
components, with which they all together amount to $N^2-1$ vectors.
This provides a means to expand any adjoint color vector in this basis,
with the expansion coefficients being gauge invariant quantities.  We
also need to define appropriate gauge invariant variables.  These should
be $2(N^2-1)$ in number, and will depend on the invariant tensors of
$SU(N)$.  For $SU(3)$, in particular, they are
\eqn\gvar{\eqalign{g_{ij}=&u^a_iu^a_j\cr h_{ijk}=&d^{abc}u^a_iu^b_j
u^c_k\ ,}} where $d^{abc}$ is the totally symmetric symbol of $SU(3)$.
These represent precisely the $16=6+10$ gauge invariant degrees of
freedom for $SU(3)$ at each space point.

It is now possible to explore \defun\ further for $SU(3)$ along the
lines of Sec. 3. Analogously to \covu , one can write \eqn\covun{\pa_j
u^a_k+f^{abc}A^b_j u^c_k\ =\ S^i_{jk}u^a_i+S^{\{mn\}}_{jk}u^a_{\{mn\}}\
,} where $S_{jk}=S_{kj}$ and one uses the as yet undefined basis
\basissun . We now define $u^a_{\{mn\}}$ through:
\eqn\uamn{\eqalign{u^a_iu^a_{\{mn\}}=&0\cr
p^a_{\{ij\}}u^a_{\{mn\}}=&g_{\{ij\}\{mn\}}={1\over 2}(g_{im}g_{jn}+
g_{in}g_{jm}-{2\over 3}g_{ij}g_{mn})\ ,}} where $p^a_{\{ij\}}\equiv
d^{abc}u^b_{\{i}u^c_{j\}}$, and $\{\ldots\}$ again means symmetrizing
and removing the trace.  These are $15+25=40$ linear equations for the
40 quantities $u^a_{\{mn\}}$.

It is useful to define two 8-beins: \eqn\eightbein{\eqalign{ u^a_X=&\{
u^a_i,u^a_{\{mn\}}\}\cr p^a_X=&\{ u^a_i,p^a_{\{mn\}}\}\ ,}} where $X$
stands for $i$ or $\{mn\}$, running altogether over 8 indices.  Their
inverses are: \eqn\eightbeininv{\eqalign{ n^{aX}=&\{
u^{ai},p^{a\{mn\}}-u^a_\ell h^{\ell\{mn\}}\}\cr \tilde{n}^{aX}=&\{
u^{ai}-u^a_{\{k\ell\}}h^{i\{k \ell\}}\}\ ,}} where space indices here
and throughout are raised with the inverse metric
$g^{XY}\equiv\{g^{ij},g^{\{ij\}\{mn\}}\}$ found from \gvar\ and \uamn.
It is simple to verify that
\eqn\un{\tilde{n}^{aX}p^a_Y=n^{aX}u^a_Y=\delta^X_Y\ ,} which then
implies \eqn\nu{\tilde{n}^{aX}p^b_X=n^{aX}u^b_X=\delta^{ab}\ .} A formal
expression for $u^a_{\{mn\}}$ is now simple to obtain.  Given the matrix
\eqn\matm{M^{\{ij\}\{mn\}}\equiv n^{a\{ij\}}n^{a\{mn\}}= p^{a\{ij\}}
p^{a\{mn\}}-h_k^{\{ij\}}h^{k\{mn\}}\ ,} the quantities $u^a_{\{mn\}}$
will be: \eqn\uexpl{ u^a_{\{mn\}}=M^{-1}_{\{mn\}\{ij\}}n^{a\{ij\}}\ .}

Applying similar manipulations to those described below \covu, together
with the use of \uamn, one obtains an explicit expression for
$S^i_{jk}$: \eqn\s{S^i_{jk}=\left\{ \matrix{i \cr jk \cr}\right\} \ ,}
as in \christoffel , with the metric given by \gvar . The remainder of
\covun, {\it i.e.}, the quantity $R^a_{ij}= S^{\{mn\}}_{ij}u^a_{\{mn\}}$
can be found from studying the quantity $\nabla_ih_{jk\ell}$.  It is not
difficult to arrive at the following $30+18=48$ linear equations
defining the 48 quantities $R^a_{ij}$:
\eqn\raij{\eqalign{R^a_{ij}u^a_k=&0\cr R^a_{i(j}p^a_{k\ell
)}=&\nabla_ih_{jk\ell}\ .}} After some manipulations one finds the
following solution to the above inhomogeneous system:
\eqn\rsol{R^a_{ij}={1\over3}u^{a\{mn\}}\left(
\nabla_{(i}h_{j)mn}-{1\over2}\nabla_{(m}h_{n)ij}\right)\ .} Whether the
homogeneous system has a nontrivial solution, thus leading to an extra,
``zero mode", term in $R^a_{ij}$ above, is a question that can be
decided by computing the appropriate $48\times48$ determinant, possibly
through a symbolic manipulation program.  For the purposes of the
present work, we shall simply assume \rsol\ is the full solution for
$R^a_{ij}$.

The construction of the magnetic field in terms of geometric variables
can finally be outlined.  Like in the $SU(2)$ case, one must start by
considering the gauge Ricci identity: \eqn\riccisuth{\eqalign{
[D_i,D_j]\ u^a_k=&T^a_{ijk} =-\hat{\epsilon}_{ijm}f^{abc}B^{cm}u^b_k\cr
[D_i,D_j]\ p^a_{\{mn\}}=&S^a_{ij\{mn\}}
=-\hat{\epsilon}_{ijm}f^{abc}B^{cm}p^b_{\{mn\}} \ ,}} where $T^a_{ijk}$
and $S^a_{ij\{mn\}}$ are quantities to be calculated using \covun, \s\
and \rsol.  This will not be done here, as it is a lengthy but
straightforward exercise.  Once these quantities are calculated we are
essentially done, as contraction of the above with the 8-bein
$\tilde{n}^{aX}$ effectively isolates the $B$ field:
\eqn\bfield{\eqalign{
-\hat{\epsilon}_{ijm}f^{abc}B^{cm}p^b_X\tilde{n}^{dX}=&\cr
-\hat{\epsilon}_{ijm}f^{adc}B^{cm}=&
(T^a_{ijk}\tilde{n}^{dk}+S^a_{ij\{mn\}} \tilde{n}^{d\{mn\}} )\ .}} The
final expression for the $B$ field then is: \eqn\bfieldfinal{
B^{ai}=-{1\over6}f^{abc}\epsilon^{ijk}
(T^b_{jkm}\tilde{n}^{cm}+S^b_{jk\{mn\}} \tilde{n}^{c\{mn\}} )\ .}

This concludes the outline of the extension of the formalism to larger
gauge groups, and in particular $SU(3)$.  Naturally, the lengthy task
remains of calculating the electric field also in the fashion presented
above for the magnetic field.  We have not done this here, as we believe
it is presently more important to pursue further our formalism for gauge
group $SU(2)$.\bigskip

\newsec{Conclusions}\bigskip

It has been shown that the Hamiltonian canonical variables of Yang-Mills
field theory naturally lend themselves to the implementation of general
coordinate transformations where the variables transform as $GL(3)$
tensors.  This is true for any spatial metric.  With the introduction of
a special metric, which is defined so that the Yang-Mills vector
potential is the spin-connection for that metric, an ordinary Riemannian
geometry results.  Although the Yang-Mills Hamiltonian is not a
geometrical invariant, the Gauss law constraint on states is easy to
enforce in terms of the geometrical variables.  It is possible to show
that in fact gauge invariant states must be functions only of these
gauge invariant metric variables.  Another important consequence of the
Gauss law constraint is that gauge invariant states with finite color
electric fields must obey further gauge invariant constraints.  Thus,
Gauss' law is seen to enforce conditions beyond simply those of
invariance under infinitesimal gauge transformations.  This bears a
resemblance with the energy barriers found in other gauge invariant
geometric approaches to Yang-Mills theory \fhjl\bfh.

The case of the $SU(2)$ gauge theory has been fully worked out and the
first steps have also been given for the $SU(>2)$ fields.  In terms of
the metric theory a closely related $GL(3)$ invariant field theory can
also be defined.  These results are completely formal and no discussion
of the effects of renormalization has been given.  The consequences of
the necessary introduction of a cutoff and the practical utility of this
reformulation await future study.

\bigskip\noindent {\bf Acknowledgments}\bigskip

We would like to thank our colleagues at MIT and at the University of
Barcelona for many discussions, and in particular D.Z.  Freedman for his
constructive criticism and constant interest.

\bigskip \appendix{} \bigskip

The existence of a well-defined inverse ${\cal H}$ of $T$ (cf. \green )
depends upon the nature and number of zero modes of $T$.  It is obvious
that there is at least one such zero eigenvalue eigenfunction since the
metric tensor $g^{ij}$ obeys $\nabla_kg^{ij}=0$.  One may establish a
relationship between the zero modes of $T$ and multiple solutions of the
defining relation between the vector potential and the dreibein $u$,
\defu.  It has already been remarked that \defu\ defines a relation
between $u$ and $A$ up to the global scale globally homogeneous in $g$.
If the operator $T$ is to have an inverse in the space of functionals of
$g$ when $T$ has a zero mode eigenfunction $z_{ij}$, then
\eqn\globalconst{\int d^3\!x\ z_{ij}(x) {\d\Psi\over\d g_{ij}(x)}=0\ .}
This follows from \aiai\ by mutiplication by $z_{ij}$ and integrating
the right side by parts.  With $z_{ij}=g_{ij}$, we find
\eqn\scaleinv{\int d^3\!x\ g_{ij}(x) {\d\Psi\over\d g_{ij}(x)}=0\ ,}
which is precisely the condition that $\Psi$ is globally homogeneous in
$g$.  One may see directly that if in addition there are other zero mode
eigenfunctions of the operator $$ \epsilon^{imj}D^{ab}_m\ ,$$ since it
appears in the $\d A/\d u$ term of the chain rule \eqn\chrule{
{\d\over\d g_{ij}(x)} = \int d^3\!y d^3\!z\ {\d u^b_m(z)\over\d
g_{ij}(x)}\ {\d A^a_k(y)\over\d u^b_m(z)}\ {\d\over\d A^a_k(y)} \ ,} its
singularities coincide with those of $T$.  Indeed the zero mode
eigenfunctions $w$ and $z$ (of $\epsilon D$ and $T$ respectively) are
related by \eqn\zerom{w^a_i = u^{am}z_{im}\ .} Thus there is a
not too surprising parallel between this ambiguity in the vector
potential and the one associated with the electric field.

However, this problem is made somewhat clearer if it is recognized that
``zero" is a very special place in the spectrum of eigenvalues of these
operators.  It is clear from \top\ that eigenfunctions which have
asymptotically large covariant curls will be associated with eigenvalues
which are asymptotically large.  At the same time it follows from the
covariant divergence of \top ,
\eqn\div{\nabla_i\phi_\a^{ij}={1\over\l_\a}{\epsilon^j_{\ mn}
\over\sqrt{g}}\ G^m_{\ k}\ \phi_\a^{kn}\ ,} that an asymptotically large
covariant divergence eigenfunction will be associated with eigenvalues
where $1/\l_\a$ becomes asymptotically large.  Thus there must be a
continuous spectrum of eigenvalues with accumulation points at both zero
and infinity.  Since the operator has no well-defined sign these
eigenvalues must be of both algebraic signs.  Thus, in a finite volume
all of the eigenfunctions will be associated with a $1/\sqrt{V}$
normalization.  Any finite ($<V$) number of exactly zero eigenfunctions,
will then be of no consequence in the limit of infinite volume.
However, care will have to be taken in constructing the Green's function
which appears in \green\ so that the formal sum over the spectrum which
asymptotes to zero yields a distribution which as a consequence of the
covariant divergence obeys the formal constraint,
\eqn\graddel{{\epsilon^j_{\ mn}\over\sqrt{g}}\ G^m_{\ p}\ {\cal
H}^{pn}_{\ \ k\ell}={1\over\sqrt{g}}\ \d^j_\ell\nabla_k\d (x-y)\ .}

\listrefs \end